%% file: paper.tex
\documentclass[runningheads]{llncs}
\usepackage[T1]{fontenc}
\usepackage{float} 
\usepackage{listings}
\usepackage{amsmath}
\usepackage{siunitx}
\usepackage{hyperref}
\usepackage{placeins}
\usepackage{xcolor}
\usepackage{graphicx}
\usepackage{caption}
\usepackage{subcaption}
\usepackage{color}
\input{setup/color.tex}

\input{setup/lstlisting.tex}
\begin{document}
\title{\textsc{Madvex}: Instrumentation-based Adversarial Attacks on Machine Learning Malware Detection}
\titlerunning{Madvex: Adversarial Attacks on ML Malware Detectors}
\author{Nils Loose \and Felix Mächtle \and Claudius Pott \and Volodymyr Bezsmertnyi \and Thomas Eisenbarth }
\authorrunning{N. Loose \emph{et al.}}
\institute{Institute for IT-Security, Ratzeburger Allee 160, 23562 Lübeck, Germany\\
\email{\{n.loose, f.maechtle, c.pott, thomas.eisenbarth\}@uni-luebeck.de, volodymyr.bezsmertnyi@student.uni-luebeck.de}}
\maketitle
\def\chapterautorefname{Chapter}%
\def\sectionautorefname{Section}%
\def\subsectionautorefname{Subsection}%
\def\subsubsectionautorefname{Subsubsection}%
\def\paragraphautorefname{Paragraph}%
\def\tableautorefname{Table}%
\def\equationautorefname{Equation}%
% --- Abstract ---
\begin{abstract}
\input{chapter/Abstract.tex}
\keywords{Malware Detection  \and Adversarial Attack \and Binary Instrumentation \and \textsc{Minos} \and Cryptojacking}
\end{abstract}
% --- Chapters ---
\input{chapter/Introduction.tex}
\input{chapter/Background.tex}
\input{chapter/Main-Chapter.tex}
\input{chapter/Evaluation.tex}
\input{chapter/Related-Work.tex}
\input{chapter/Conclusion.tex}
% --- Acknowledgements ---
\subsubsection{Acknowledgements.} 
We thank the reviewers and our shepherd for their helpful comments and suggestions.
This work has been supported by ERDF through the EMSIK project and by BMBF through the PeT-HMR project.
% --- Bibliography ---
\bibliographystyle{splncs04}
\bibliography{paper}
\end{document}

%% file: setup/color.tex
\definecolor{uzlmain}{RGB}{0,75,90}
\definecolor{uzlpal1}{RGB}{232,192,173}
\definecolor{uzlpal1a}{RGB}{228,32,50}
\definecolor{uzlpal1b}{RGB}{181,22,33}
\definecolor{uzlpal1c}{RGB}{127,21,24}
\definecolor{uzlpal2}{RGB}{240,205,178}
\definecolor{uzlpal2a}{RGB}{236,116,4}
\definecolor{uzlpal2b}{RGB}{203,81,25}
\definecolor{uzlpal2c}{RGB}{129,53,19}
\definecolor{uzlpal3} {RGB}{236,218,136}
\definecolor{uzlpal3a}{RGB}{250,187,0}
\definecolor{uzlpal3b}{RGB}{191,134,20}
\definecolor{uzlpal3c}{RGB}{117,83,17}
\definecolor{uzlpal4} {RGB}{219,215,187}
\definecolor{uzlpal4a}{RGB}{149,188,14}
\definecolor{uzlpal4b}{RGB}{126,133,37}
\definecolor{uzlpal4c}{RGB}{50,88,75}
\definecolor{uzlpal5} {RGB}{205,219,216}
\definecolor{uzlgreenlight}{RGB}{59,178,160}
\definecolor{uzlgreen}{RGB}{36,143,133}
\definecolor{uzlgreendark}{RGB}{0,90,91}
\definecolor{uzlpal6}{RGB}{198,220,226}
\definecolor{uzlmainlight}{RGB}{0,174,199}
\definecolor{uzlpal6b}{RGB}{0,145,168}
\definecolor{uzlpal6c}{RGB}{0,97,122}
\definecolor{uzlpal7}{RGB}{194,217,230}
\definecolor{uzlpal7a}{RGB}{60,169,213}
\definecolor{uzlpal7b}{RGB}{0,131,173}
\definecolor{uzlpal7c}{RGB}{0,88,119}
\definecolor{uzlpal8}{RGB}{195,217,237}
\definecolor{uzlpal8a}{RGB}{110,165,206}
\definecolor{uzlpal8b}{RGB}{0,106,163}
\definecolor{uzlpal8c}{RGB}{0,70,114}
\definecolor{uzlgrey}{RGB}{101,101,101}

\definecolor{materialgreen}{RGB}{104,159,56}
\definecolor{materialred}{RGB}{230,74,25}

\definecolor{highlightcolor}{RGB}{138,198,205}
\definecolor{highlightcolor2}{RGB}{205,198,138}

\definecolor{materialcyan}{RGB}{0, 188, 212}
\definecolor{materialorange}{RGB}{255, 87, 34}

\definecolor{teal500}{RGB}{0, 150, 136}
\definecolor{cyan500}{RGB}{0, 188, 212}

%% file: setup/lstlisting.tex
\newcommand{\LstHighlight}[1]{\makebox[0pt][l]{\color{highlightcolor!20}\rule[-3pt]{#1\linewidth}{10pt}}}

\lstdefinelanguage{WebAssembly}{
  sensitive=true,
  otherkeywords={},
  morekeywords=[1]{i32,f32,i64,f64,local,global},
  keywordstyle={[1]\color{teal500}},
  morekeywords=[2]{1,152,51,1,49},
  keywordstyle={[2]\color{c yan500}},
  morekeywords=[3]{add,const}
  keywordstyle={[3]\color{bluemunsell}},
  morekeywords=[4]{},
  keywordstyle={[4]\color{candypink}},
  morekeywords=[5]{module, func, param, result, global, get_global, mut, set_global, export, import, memory, data, get_local, set_local, elem, table, call,call_indirect, type},
  keywordstyle={[5]\color{blue}},
  morekeywords=[6]{=,;},
  keywordstyle={[6]\color{britishracinggreen}},
  morekeywords=[7]{(,),[,],.},
  keywordstyle={[7]\color{black}},
  numberstyle=\tiny\color{black},
  rulecolor=\color{black},
  morecomment=**[l][\itshape\color{greencomments}]{;;},
}
\lstdefinestyle{wasm}
{
    frame=lines,
    language=WebAssembly, 
    basicstyle = \scriptsize\ttfamily, 
    showspaces=false, 
    showtabs=false,
    showstringspaces=false,
    breaklines=true, 
    breakatwhitespace=true,
}

%% file: chapter/Abstract.tex
WebAssembly (\texttt{Wasm}) is a low-level binary format for web applications, which has found widespread adoption due to its improved performance and compatibility with existing software. However, the popularity of \texttt{Wasm} has also led to its exploitation for malicious purposes, such as cryptojacking, where malicious actors use a victim's computing resources to mine cryptocurrencies without their consent. To counteract this threat, machine learning-based detection methods aiming to identify cryptojacking activities within \texttt{Wasm} code have emerged. 
It is well-known that neural networks are susceptible to adversarial attacks, where inputs to a classifier are perturbed with minimal changes that result in a crass misclassification. While applying changes in image classification is easy, manipulating binaries in an automated fashion to evade malware classification without changing functionality is non-trivial. In this work, we propose a new approach to include adversarial examples in the code section of binaries via instrumentation. The introduced gadgets allow for the inclusion of arbitrary bytes, enabling efficient  adversarial attacks that reliably bypass state-of-the-art  machine learning classifiers such as the CNN-based \textsc{Minos} recently proposed at NDSS 2021. We analyze the cost and reliability of instrumentation-based adversarial example generation and show that the approach works reliably at minimal size and performance overheads. 

%% file: chapter/Introduction.tex
\section{Introduction}
With the introduction of WebAssembly (\texttt{Wasm}) in 2017, web applications are able to utilize a system's CPUs with near-native efficiency \cite{WebAssemblyCoreSpecification2}. \texttt{Wasm} allows developers to make computationally heavy applications available in-browser and has since been used for games, text processing, visualizations, and media players \cite{empirical-study,musch-cryptojacking-study}. On the downside, malicious parties have also utilized \texttt{Wasm} to distribute malicious binaries to victims that visit an infected website and thus gain access to the victim's resources without having to gain access to their system. 
In particular, the near-native performance of \texttt{Wasm} and the support provided by all major browsers make WebAssembly a prime target for cryptojacking attacks \cite{empirical-study,musch-cryptojacking-study,sok_Cryptojacking_Malware}.
In-browser cryptojacking or drive-by cryptocurrency mining allows an attacker to utilize their victim's computational resources for mining cryptocurrencies without their knowledge or consent, thus profiting from the returns without having to pay for the spent energy.
To address this issue, various methods have been proposed to protect against cryptojacking attacks.
However, while fast, traditional static approaches like blacklisting malicious hosts or matching signatures are easily bypassed~\cite{72-MinerRay}.
Dynamic detection systems \cite{minesweeper,77-dynamic-svm,detection:api-ressource-monitoring}, on the other hand, rely on more sophisticated metrics that cause a runtime overhead and require the malicious binary to be executed. 
\textsc{Minos}, a lightweight machine learning-based detection system, provides a promising solution to this problem~\cite{minos}. 
By transforming \texttt{Wasm} binaries to grey-scale images, \textsc{Minos} can utilize a convolutional neural network (CNN) for the classification of binaries.
This provides a rapid and effective approach that can be applied prior to executing the binaries, thereby offering efficient protection against in-browser cryptojacking attacks.
While promising, CNNs are known to be susceptible to adversarial attacks \cite{advex_summary}. 
Malicious parties looking to distribute their malware have a high incentive to evaluate possible avenues for bypassing detection frameworks. In particular, the development of more sophisticated evasion techniques by attackers could render existing detection methods ineffective. 
Adversarial examples are usually crafted under the assumption that small changes to the input are neglectable.
However, applying adversarial examples to binaries that follow strict syntactical and semantical rules requires specific placement of adversarial payloads without invalidating the binary or changing the semantics.
Still, attacks leveraging adversarial examples to bypass visualization-based malware detectors have been proven to succeed on Windows Portable Executables \cite{12-vis-nonexec,13-COPYCAT-vis-eof-append,14-vis-nop-align}. 

In this paper, we evaluate the feasibility of utilizing adversarial examples against the \texttt{Wasm}-based classifier \textsc{Minos} \cite{minos} presented at NDSS 2021.
 We demonstrate the feasibility of inserting semantic-preserving gadgets using binary instrumentation into the code section of WebAssembly applications, allowing effective crafting of adversarial examples inside the gadget, thus enabling the evasion of the \textsc{Minos} detection system. In contrast to existing work, we add the adversarial payload directly into the application's control flow and introduce both size-efficient (SE) and optimization-resistant (OR) gadgets. Our findings shed light on the potential weaknesses of machine learning-based classifiers in detecting cryptojacking and highlight the need for ongoing efforts to improve their robustness and security, particularly when classifiers are applied in scenarios with incentives to evade classification.
To summarize, our key contributions are:
\begin{itemize}
\item Comprehensive collection of malign \texttt{Wasm} samples from the \textit{Cisco Umbrella 1 Million} websites list.
\item A novel approach for automatically crafting adversarial examples in code by introducing semantic-preserving instruction gadgets via instrumentation.
\item Demonstrating a grey-box adversarial attack against the \textsc{Minos} classifier by training a substitute model and applying our gadgets.
\item A comprehensive evaluation of the efficacy and costs of the attack.
\end{itemize}

%% file: chapter/Background.tex
\section{Background}
\label{sec:background}
\subsection{WebAssembly} 
\label{sec:background:wasm}
WebAssembly (\texttt{Wasm})~\cite{WebAssemblyCoreSpecification2} is a binary instruction format for a stack-based virtual machine that enables high-performance applications that run seamlessly in web browsers. It is designed to provide near-native performance to web applications and allows developers to write applications in various programming languages, including \texttt{C}, \texttt{C++}, and Rust, while still being executed in the browser. 
\texttt{Wasm} is supported by all major web browsers and has gained significant traction in recent years, particularly in resource-intensive applications, where the performance benefits provided by \texttt{Wasm} are especially important.
In most settings, \texttt{Wasm} is integrated into the JavaScript code of a website, from where the \texttt{Wasm} modules are loaded, and the respective functions are called.
Its stack-based architecture, widespread support, and versatility make it an essential tool for modern web development.
\subsection{Cryptojacking Malware}
\label{sec:background:Cryptojacking}
\textit{Cryptocurrency mining} is the process of solving complex mathematical problems in order to validate transactions and add new blocks to a blockchain network~\cite{Originales_Bitcoin_Paper}. The process requires a significant amount of computational power and energy.
As compensation for the computation time, miners are rewarded with new units of the respective cryptocurrency. This reward mechanism is a key component of the decentralized nature of many cryptocurrencies, as it incentivizes individuals and organizations to participate in the network and maintain its security. However, as the difficulty of mining increases and the competition among miners grows, the margin between the resources spent on mining and the returned profits diminishes.
If a malicious actor manages to utilize a victim's resources for mining, the computational cost is removed from the equation.
In general, the unauthorized use of a device's computing power to mine cryptocurrencies, typically without the knowledge or consent of the device's owner, is referred to as \textit{cryptojacking}. This type of attack can occur via host- or browser-based mining and can have significant impacts on both individual users and organizations. 
Host-based cryptojacking requires the installation of a cryptocurrency miner on the victim's machine through, i.e., malicious software installed by the victim~\cite{sok_Cryptojacking_Malware}.
Browser-based cryptojacking is a method of exploiting a victim's device through a malicious website. The attacker inserts a script into the website's code that runs in the victims' browser upon visiting the site and uses their device's processing power to mine cryptocurrency while profiting the owner of the operation. 
With the introduction of WebAssembly and its near-native speed, the efficiency of browser-based mining has significantly increased, making the attack lucrative.
Unlike traditional malware, browser-based cryptojacking does not require the victims to download any files, making it subtle and difficult to prevent.
\subsection{Malware Detection}
\label{sec:background:Malware detection}
\label{sec:background:minos}
Identifying whether a binary contains malicious functionality is an active area of research across different types of binaries.
Various approaches have been proposed for detecting \textit{cryptojacking}, one of the primary malicious usages of \texttt{Wasm} binaries~\cite{musch-cryptojacking-study}.
Due to the reliance of cryptojacking malware on network communication, network-based detection systems have been proposed, analysing the network traffic \cite{network-based-detection}. 
Host-based detection frameworks rely, in general, on either static or dynamic analysis to identify malware.
Dynamic approaches observe the execution of a binary while monitoring key metrics such as memory consumption~\cite{75-dynamic-mem-consumption}, the number of executed arithmetic operations~\cite{SEISMIC}, or through CPU profiling~\cite{minesweeper}. 
Prevention techniques that identify malware based on resource consumption can be circumvented through throttling~\cite{73-static-hash-detection}.
Additionally, a number of machine learning classifiers have been proposed that require dynamic features such as API calls and resource information~\cite{detection:api-ressource-monitoring} or runtime information such as the number of web sockets or workers~\cite{77-dynamic-svm}.
In order to generate dynamic features, the potentially malicious binary needs to be executed on the host's machine. 
Static approaches, on the other hand, do not require the evaluated code to be executed; instead, the binary is directly evaluated, for example, by matching known signatures or URL blacklisting~\cite{73-static-hash-detection}. However, these techniques can be circumvented using obfuscation~\cite{72-MinerRay}. 
MinerRay~\cite{72-MinerRay} relies on the static detection of hash semantics to make obfuscation-based prevention harder as the semantics of the functions are evaluated.

In general, efficiently detecting whether a WebAssembly binary utilizes the host's resources for mining cryptocurrencies without relying on dynamic features allows a detection framework to warn the user that a malicious binary is loaded before the execution of the binary. 
Nassem \emph{et al.} developed \textsc{Minos}~\cite{minos}, a lightweight real-time detection system that aims to efficiently detect whether a WebAssembly binary utilizes the host's resources for cryptomining using a CNN.
\textsc{Minos} is designed to be implemented as a browser plug-in which uses the detection framework to warn users about any detected cryptomining binaries before they are executed.
Upon visiting a website that loads a \texttt{Wasm} binary, the detection framework transforms the bytes contained inside the binary into a two-dimensional grey-scale image which is then evaluated by a pre-trained CNN. This architecture allows the system to classify a binary, on average, in $25.9\; ms$ while achieving an overall accuracy of $98.97\%$ against an in-the-wild dataset \cite{minos}.
\subsection{Adversarial Attacks}
\label{sec:background:advattacks}
Deep neural networks, along with other machine learning models, have been discovered to be susceptible to adversarial attacks on their input data \cite{DNN-are-susceptible-to-adv-att,evasion-attack-test-time}. 
Given a target model $\theta$, an input $x$ and a target class $t\neq \theta(x)$, an adversaries objective is to find a minimal perturbation $\delta_x$ under a norm $\mathcal{N}=||~\cdot~||$ s.t.  
\begin{equation}\label{eq:adv}
\theta(x+\delta_x) = t
\end{equation}
Minimizing the perturbation vector $\delta_x$ under a norm $\mathcal{N}$ ensures that the original input $x$ and the newly generated input, or \textit{adversarial example}, $x^*=x + \delta_x$ are close to each other under a given distance metric $\mathcal{D}$. 
However, finding a perturbation $\delta_x$ that satisfies \autoref{eq:adv} is generally a hard problem due to the nonlinearity of the evaluated model $\theta$ \cite{DNN-are-susceptible-to-adv-att}. 
Existing methods for crafting an adversarial example, such as the L-BFGS, solve the problem using approximations~\cite{advex_summary}.
Carlini and Wagner (C\&W) proposed a different approach by transforming the constraint shown in \autoref{eq:adv} into an optimization problem using an appropriately chosen objective function $\mathcal{L}$, s.t. if $\theta(x+\delta_x) = t$ is satisfied, $\mathcal{L}(x^*)\leq0$ holds  \cite{Carlini_Wagner_Adv_example}. 
By moving the constraint into the minimization term, the problem of finding an adversarial example is an optimization task that minimizes $\mathcal{N}(\delta_x) + \epsilon \cdot \mathcal{L}(x^*)$ such that $x^* \in [0,1]^n$ where $\epsilon>0$ is a suitably chosen constant.
The optimization problem is solved using gradient-based optimization methods \cite{DBLP:Cross_Entropy_loss}. 
The gradient of the objective function with respect to the input $x$ is used to update the perturbation $\delta_x$ in each iteration of the optimization process.
The process is repeated until the minimum perturbation, which results in the adversarial example being classified as the target class t, is found. 
Without access to the gradients of the target model $\theta$, the aforementioned attack cannot be utilized.
However, given query access, the adversary can train a local substitute network \cite{black-box-adv-attck} by querying the target classifier with synthesized or otherwise gathered data. Using the results obtained through inference against the target network as labels, the local model is trained. 
Due to the \textit{transferability} between models, it is possible to train a machine learning model that mimics the behaviour of a target model~\cite{explaining_advex}. In a black-box scenario \cite{black-box-adv-attck}, a network with unknown architecture is attacked, requiring a custom architecture for the local substitute network. 
In the grey-box scenario, additional information about the target network, such as parameters or its architecture, is known, and hence the substitution network architecture can be chosen similarly to the target model. 
The local model can then be utilized to generate adversarial examples that are transferable to the target network \cite{black-box-adv-attck}.

%% file: chapter/Main-Chapter.tex
\section{\textsc{Madvex}: Crafting Functional Adversarial Binaries}
\input{figures/Data-Preparation-Pipeline.tex}%
\label{sec:main}
The \textsc{Minos} classifier \cite{minos} uses an image-based machine learning technique to quickly identify malicious WebAssembly binaries. However, such classifiers are shown to be vulnerable to adversarial attacks \cite{DNN-are-susceptible-to-adv-att}. This section describes the attack methodology used to craft binaries that are misclassified by \textsc{Minos}. To illustrate the applicability of such an attack, we limit the adversary and assume a grey-box scenario where the attacker has query access to the model and knowledge of the network's architecture. Although the Minos classifier's architecture was published by Naseem \emph{et al.}, the training data and model were not made available. Therefore, we use a \textsc{Minos} classifier trained by Cabrera-Arteaga \emph{et al.}\ \cite{wasm-evade} as the target of our attack experiments. 
\subsection{Data Acquisition}
\label{sec:main:dataacquisition}
The performance of the attack correlates with the quality of the local substitute model trained by the adversary. Therefore a comprehensive dataset of malicious and benign WebAssembly binaries is required to train a suitable substitute network. The original \textsc{Minos} model was trained on a balanced dataset containing 300 samples \cite{minos}.
The data preparation and training procedure for the substitute model is schematically visualized in Fig. \ref{fig:data-prep} and described below in detail. 
To obtain benign samples, we used WasmBench\footnote{\url{https://github.com/sola-st/WasmBench} (Accessed 2023/01/31)}, a WebAssembly dataset containing more than $23.000$ real-world binaries published by Hilbig \emph{et al.}\ as part of an empirical study \cite{empirical-study}. 
We obtained $34$ malicious samples from a dataset\footnote{\url{https://github.com/vusec/minesweeper} (Accessed 2023/01/31)} published in the context of Minesweeper \cite{minesweeper}.
Additionally, we ran a crawler to increase the number of malware samples and gather up-to-date malware. By iterating over the Cisco Umbrella $1$ Million list \cite{cisco-1mil}, we were able to download $187$ WebAssembly binaries. Each domain on this list is visited by the crawler, which resides on any page for three seconds. By hooking a JavaScript function into each document load, we are able to dump any WebAssembly binary before it is executed. Considering that the malware may not reside on the homepage directly, the crawler additionally visits three randomly chosen internal links. Overall $40\%$ of the crawled binaries resided on subdomains and were found either through accessing internal links or redirects. 
The Minesweeper \cite{minesweeper} classifier categorized ten out of the $187$ crawled binaries as being malicious.
Even after combining the samples of public datasets with the results of our crawling campaign, the number of obtained malicious binaries is considerably lower than that of benign binaries. In order to compensate for this difference and additionally increase the number of samples, we utilize the \texttt{Wasm}-fuzzer \texttt{wasm-mutate} \cite{wasm-mutate} as a diversifier. By utilizing \texttt{wasm-mutate}, one can generate a variety of different WebAssembly binaries that retain the original semantic. % while differing in various aspects. 
Mutation cores available in \texttt{wasm-mutate} enable semantic-preserving transformations.
A sample function that performs the addition of two integers and two mutations of the function are shown in Fig. \ref{fig:wasm-mutate}. 
Each mutation is generated using a different seed, allowing us to generate a larger variety of syntactically different binaries with identical semantics. 
To generate appropriate adversarial examples, a shadow model that is as similar to the target model as possible must be utilized. To achieve this, the internal labels assigned to the samples are only used for balancing and not used for training. Instead, the pre-trained \textsc{Minos} network \cite{wasm-evade} is employed for label generation. After augmentation of the malicious samples, we obtain a dataset containing $\num{2.3e4}$ malicious and $\num{2.3e4}$ benign binaries that are used for training the substitute model. 
\input{figures/Mutations.tex}%
\subsection{Substitute Network Training}
\label{sec:main:train-substitute-model}
We use the architecture employed by \textsc{Minos} for the substitute model because we assume a known architecture in the grey-box attack. The architecture of the CNN is shown in Fig. \ref{fig:minos-architecture}. 
Convolutional neural networks typically receive an image as the input for classification. The \textsc{Minos} classifier requires the input to be a grey-scale image of size $100\times100$. 
To allow binaries of varying sizes to be represented as a fixed-dimensional image, the bytes are reshaped into the largest possible two-dimensional array with the same width and height. The remaining bytes are discarded. Initially, each byte of the binary corresponds to one pixel. However, the image is downscaled to a $100\times100$ image. A detailed description of the downsampling process is given in Section \ref{sec:main:downsampling}.
\input{figures/Minos-Architecture.tex}
The original model was trained using an $80\%$ training and $20\%$ testing split. However, we use $5$-fold cross-validation for training. Hence five models are trained each on $80\%$ of the dataset described in Section \ref{sec:main:dataacquisition}, while $20\%$ of samples are withheld for validation. 
For the evaluation, \textsc{Minos} was trained with one epoch (M-1) to prevent overfitting, followed by 50 epochs (M-50), the same number as the target model. The area under the curve (AUC) and loss after the final epochs are reported in Table \ref{tab:training}. Even after training the substitute network for only one epoch, the validation AUC reaches $99\%$ with a validation loss of $0.14$. After training for $50$ epochs, the validation loss decreases to $0.04$.
\input{figures/Training-Stats}%
\subsection{Attack Methodology}
\label{sec:main:adversarial-attack}
\begin{figure}[t]
    \centering
    \includegraphics[width=\textwidth]{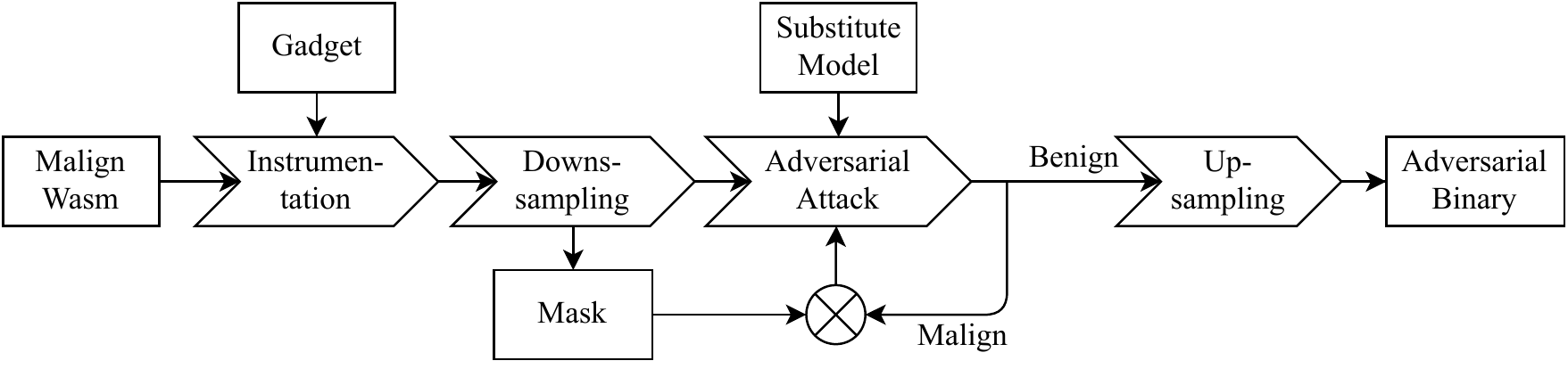}
    \caption{Schematic overview of the attack methodology. A malicious binary is instrumented to add the gadgets used for carrying the adversarial payload. After downsampling, the adversarial attack is performed against the substitute model. To recreate the original binary, we upsample the adversarial image and recreate the original binary.}
    \label{fig:adv-ex-generation}
\end{figure}

\input{figures/Section-Distribution.tex}
Performing an adversarial attack against an image-based classifier requires slight modifications of the original image to manipulate the generated response in the desired direction. The alterations are often transparent to the naked eye as they result in a small amount of noise added to the original image. However, in the case of binaries, slightly manipulating the value of a pixel, for example, changing a value from $0x2A$ to $0x2B$, changes the original instruction from \texttt{f32.load} to \texttt{f64.load} invalidating the binary. We require a procedure that allows us to manipulate certain areas of the binary without changing the behaviour. 
Using instrumentation, we can add, manipulate or remove instructions from the malware and provide areas inside the code section that can be utilized for the adversarial attack. While we are still unable to manipulate arbitrary pixels, adding specially crafted gadgets into the binary enables specific bytes to be utilized for the adversarial attack.
Generating an adversarial example requires iterative manipulation of the target value in small increments. Hence, an area of bytes that are arbitrarily manipulable is ideal.
Each WebAssembly binary is split into several sections, each with a different purpose. As shown in Fig. \ref{fig:section-distribution}a, the code section represents, in most cases, the largest section inside both malicious and benign binaries that were analyzed. 
When separately evaluating the section distribution for malicious and benign binaries (cf. Fig. \ref{fig:section-distribution}b and Fig. \ref{fig:section-distribution}c), it is apparent that in both cases, the code section remains the largest section.
The code section contains all functions with their instructions, whereas the data section represents a linear array of memory accessible through instructions in the code section. 
While an attack against the data section is also possible by extending the size of the linear memory and using this area for crafting the attack, we chose to target the code section as it represents the largest section of the binaries. 
An overview of our attack methodology is given in Fig. \ref{fig:adv-ex-generation}. Each step is described in detail below.
\subsubsection{Semantic-preserving Gadgets} 
To enable manipulation inside the code section, we require an instruction that has a number of bytes that are freely choosable. In particular, instructions that load constants onto the stack cause specific values to be present inside the code section. Hence, constructing a gadget that loads an arbitrary constant onto the stack and removes it allows a number of bytes to be arbitrarily chosen. Additionally, it can be inserted anywhere into the control flow because, after the gadget's execution, the stack will be in the same state as before.
WebAssembly allows four number types to be pushed onto the stack as constants - $32$ and $64$ bit variants of integers and floats. 
We opt to use $64$ bit constants, as the ratio between the number of bytes that are available for the adversarial attack and the number of bytes required for the overall gadget is higher.
Generally, both integers and floats can work. However, WebAssembly encodes all integers using the \textit{LEB128} variable-length encoding in either the signed or unsigned variant. Compared to the encoding utilized for floating point values, \textit{IEEE-754} \cite{ieee-754}, the integer encoding enforces a number of restrictions on the bytes representing the integer. \textit{IEEE-754}, on the other hand, allows all bytes to assume all possible values. Hence we use $64$ bit floating point constants to craft the attack. 
The \texttt{f64.const x:f64} instruction can be used to push the $64$ bit floating point number \texttt{x} onto the stack. We initialize the constant to $0x80808080$ to allow both positive and negative perturbations.
To ensure that the functionality of the target binary is not modified, the value must be removed from the stack before normal execution resumes.
We demonstrate two gadgets that can be inserted after arbitrary instructions, as the execution of the gadget only changes the contents of the stack temporarily.
A \textit{size-efficient gadget} (SE) is shown in Fig. \ref{fig:gadgets}a. After the constant is pushed onto the stack, it is immediately removed again using the \texttt{drop} instruction. Each inserted gadget of this type increases the size of the binary by ten bytes, out of which the adversarial attack can utilize eight bytes (compare Fig. \ref{fig:gadgets}b). Hence, only $20\%$ of the size overhead is attributed to bytes that cannot be manipulated during the attack phase.
\input{figures/Gadget.tex}

Due to the low complexity of the size-efficient gadget, it is easy to discern that the two instructions will retain the program's semantics. However, optimizers such as \texttt{wasm-opt} \cite{wasm-opt} can remove all gadgets of this type from the binary. Note that using an optimizer before classifying the binary is not part of the \textsc{Minos} framework \cite{minos} because it would counteract the high efficiency of the detection system.  
Nevertheless, we are able to craft a gadget that is not removed by \texttt{wasm-opt}, even when using its most aggressive optimization setting. 
This resilience, however, is only made possible by increasing the gadget's complexity.
The composition of our \textit{optimizer-resistant gadget} (OR) is shown in Fig. \ref{fig:gadgets}c and the binary representation in Fig. \ref{fig:gadgets}d. The basic idea remains unchanged; we still load a constant onto the stack, thus introducing a value that can be manipulated during the attack phase. However, instead of directly loading the value onto the stack and dropping it, we use it as the increasing constant for a loop counter. However, as the value can be an arbitrary float value, i.e. negative and positive, we divide it by itself to have a known value, i.e. one. We then check whether this new value is less than some constant, i.e. $42$, which is always true, and break the loop. While it is intuitively understandable that this loop will never be executed more than once, it is not easily determined by an algorithm since loops are difficult to analyze. 
While this gadget survives optimization passes, only eight out of $32$ bytes can be utilized for the adversarial attack.
Gadgets are inserted into the code section at randomly drawn insertion points with a predetermined frequency. The relation between the number of inserted gadgets and the success rate of the attack is evaluated in Section \ref{subsec:gadget_effectiveness}.
In Section \ref{subsec:exec-speed}, we evaluate the execution speed of both gadgets in relation to the number of gadgets inserted into the binary. 
Insertion of either gadget into the target binary can be performed once per binary before distribution and requires linear time in the size of the binary, making the instrumentation efficient.
\subsubsection{Downsampling}
\label{sec:main:downsampling}
A given binary can be of any size between a few kilobytes and many megabytes. Hence, the authors of \textsc{Minos} \cite{minos} downsample each binary into an image of fixed dimensionality, i.e. $100 \times 100$ pixels (Fig. \ref{fig:minos-architecture}). As our shadow model utilizes the same architecture, it also requires an input image of that size. However, as we need to keep track of the positions that allow for a change within the instrumented binary, i.e. the constants within our gadgets, we use a custom downsampling algorithm for crafting the attacks. Yet, at inference time, the original downsampling method is used. At first, we transform the sequence of bytes $b$ from the binary into a squared image with a dimension of $\lfloor \sqrt{|b|} \rfloor $. Hence, a few bytes at the end are discarded. From this squared image, we combine as many pixels as needed in order to downsample the image to $100 \times 100$ pixels. For this purpose, we calculate the mean of a group of pixels, which then become a single pixel. To keep track of what pixels contain a byte that is used for the adversarial attack, we maintain a mask $M_1$. The mask has the same dimensionality as the image and marks all positions that contain editable values. To easily revert the downsampling when restoring the binary, we store the coordinates of the original group of pixels for each downsampled pixel. 
\subsubsection{Adversarial Attack}
After downsampling, the image $x$ is perturbed iteratively until our shadow model misclassifies the image as benign using the method proposed by Carlini \& Wagner \cite{Carlini_Wagner_Adv_example}. However, instead of optimizing for a fixed number of iterations, we keep iterating until the shadow model prediction reaches a threshold $\tau$. Experimentally we determined $\tau=10^{-13}$.
However, we also terminate the optimization after $\num{1e4}$ iterations. During our experiments, we found that the lower the threshold for the prediction score is, the higher the chance that an original model will share the classification of the shadow model.
In order to only perturb pixels related to the gadgets, we multiply the mask $M_1$ that was saved during downsampling before adding the perturbation $\delta_x$ to the sample. Given the model $\theta$, a normalization $|\cdot|$ and the constant $\epsilon$, the perturbation of the input under the objective function $\mathcal{L}$ is given as:
\[
    x = x + M_1 \cdot\epsilon\cdot \left |  \frac{\mathrm{d}}{\mathrm{d}x} \mathcal{L}(\theta(x), 0)  \right |
\]
In our experiments, we chose $\epsilon = 0.05$ and $\mathcal{L}$ as binary cross-entropy \cite{DBLP:Cross_Entropy_loss}.
We derive the change needed for the input $x$ within the normalization term so that the prediction $\theta(x)$ gets closer to zero, i.e. benign. However, instead of adding the whole perturbation to $x$, only a small factor is added. This can be compared to the learning rate in classical machine learning. As we cannot perturb the whole input image but rather just the constants within the gadgets, our crafted mask is multiplied before the summation. As the mask has zeros on all non-editable pixels, i.e. the original code of the binary, and a one wherever there is at least a single gadget, the perturbation is only applied to pixels that relate to gadgets.
\subsubsection{Upsampling}
The result of the adversarial attack is a perturbed image $x^*$ where the perturbation is only applied to the pixels that initially belonged to at least a single gadget. Those changes must now be mapped back to the original binary. 
For the perturbed image, we look at every pixel that belonged to at least one gadget. If such a pixel is found, we retrieve the corresponding group of pixels $\mathcal{G}$. To correctly update $\mathcal{G}$, the bytes belonging to an adversarial payload need to be modified s.t. the mean value of $\mathcal{G}$ equals the corresponding pixel value of $x^*$. 
Given the sum of the pixel values $\sum_{p\in\mathcal{G}} p$, the number of pixels $|\mathcal{G}|$ and the target pixel $p^*$ the update factor $f_{adv}$ can be derived using the following equation:
\[
f_{adv} = p^* \cdot |\mathcal{G}| - \sum_{p\in\mathcal{G}} p  
\]
To apply the factor $f_{adv}$ to the adversarial payload, we create a mask $M_2$ that has a one at every editable position within $\mathcal{G}$. $\overline{M_2}$ contains the same values as $M_2$ but flipped, s.t., ones become zeros and vice versa. 
We can update the group of pixels using the following equation:
\[
\mathcal{G}_{adv} = 
\begin{cases}
    M_2 \cfrac{f_{adv}}{\sum M_2} + \overline{M_2} \mathcal{G} & \text{if } \sum M_2\geq 1\\
    \mathcal{G}              & \text{otherwise}
\end{cases}
\]
The left term of the addition in the first case replaces all the editable pixels within the image with a shared factor. The second term adds the original values. This way, the new mean value of $\mathcal{G}_{adv}$ equals the target value of the downsampled image. In case there are no gadgets in the particular group, i.e. $\sum M_2 = 0$, $\mathcal{G}$ is simply copied.
After the termination of the adversarial attack, the image is flattened into a byte array $b_{adv}$, and the bytes that were cropped during downsampling are appended again.
\subsubsection{Possible Countermeasures} 
In Section \ref{subsec:gadget_effectiveness}, we show that \textsc{Minos} \cite{minos} is susceptible to the presented adversarial attack. However, it is essential to also discuss possible improvements that could prevent such adversarial attacks and aid in hardening the detection framework. 
The option to remove semantic-preserving gadgets using an optimizer was already discussed in Section \ref{sec:main}. While an additional optimization step prevents an  adversary from relying on the size-efficient gadget, the more complex optimization-resistant gadget still allows effective adversarial attacks.
Machine learning models can be directly hardened against adversarial attacks using, for example, defensive distillation \cite{Papernot2016}, which is a technique where the class probability vectors of a trained DNN are used to train another DNN of the same dimensionality. As the name suggests, defensive distillation is derived from the concept of distillation \cite{NIPS2014_ea8fcd92}, where one trained DNN is used to train a smaller DNN without losing accuracy.
Another promising method for hardening models against adversarial attacks is presented by Goodfellow \emph{et al.}\ \cite{explaining_advex}. They create adversarial examples and use them as training data for their model. 
However, the presented countermeasures were shown not to be effective against a thoughtful attacker \cite{DBLP:there_is_no_defense_against_adv_exmaples}. 

%% file: figures/Data-Preparation-Pipeline.tex
\begin{figure}[t]
    \centering
    \includegraphics[width=\textwidth]{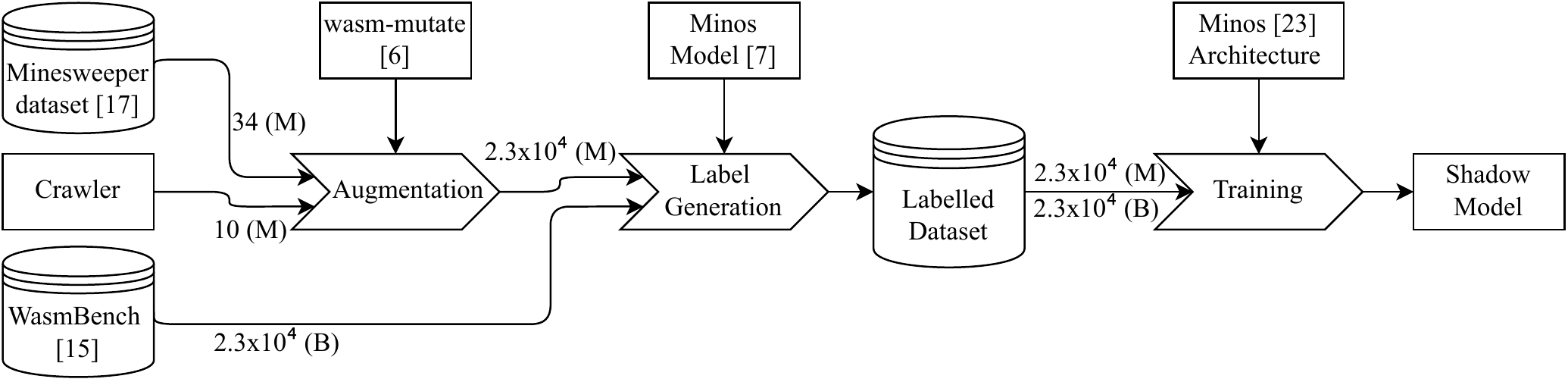}
    \caption{Systematic overview of the training procedure for the substitute model. Malicious (M) samples are augmented to generate a balanced dataset. To generate labels, the target model is queried. The labelled benign (B) and malicious data is used to train the substitute model using $5$-fold cross-validation.}
    \label{fig:data-prep}
\end{figure}

%% file: figures/Mutations.tex
\begin{figure}[t]
\centering
     \begin{subfigure}[b]{0.3\textwidth}
         \centering
         \begin{lstlisting}[style=wasm,escapechar=|,title=(a) Original function]
(func (;0;) 
  (param i32 i32) 
  (result i32)
  local.get 0
  local.get 1
  i32.add
)
\end{lstlisting}
         \label{subfig:original-function}
     \end{subfigure}
     \hfill
          \begin{subfigure}[b]{0.3\textwidth}
         \centering
         \begin{lstlisting}[style=wasm,escapechar=|,title=(b) Mutated version of (a)]
(func (;0;) 
  (param i32 i32) 
  (result i32)
  local.get 0
  i32.const 0
  i32.shl
  i32.const 0
  i32.add
  local.get 1
  i32.add
)
\end{lstlisting}
         \label{subfig:mutation-1}
     \end{subfigure}
     \hfill
          \begin{subfigure}[b]{0.3\textwidth}
         \centering
         \begin{lstlisting}[style=wasm,escapechar=|,title = (c) Mutated version of (a)]
(func (;0;) 
  (param i32 i32)
  (result i32)
  local.get 0
  local.get 1
  i32.add
  i32.const 0
  i32.shr_u
  local.get 0
  local.get 1
  i32.add
  i32.const 0
  i32.sub
  i32.and
)
\end{lstlisting}
         \label{subfig:mutation-2}
     \end{subfigure}
     \hfill
     \caption{\texttt{Wasm} function performing the addition of two integers (a) and two semantic-preserving mutations (b),(c) of the original function using different seeds in \texttt{wasm-mutate} \cite{wasm-mutate}.}
    \label{fig:wasm-mutate} 
\end{figure}%

%% file: figures/Minos-Architecture.tex
\begin{figure}[t]
    \centering
    \includegraphics[width=\textwidth]{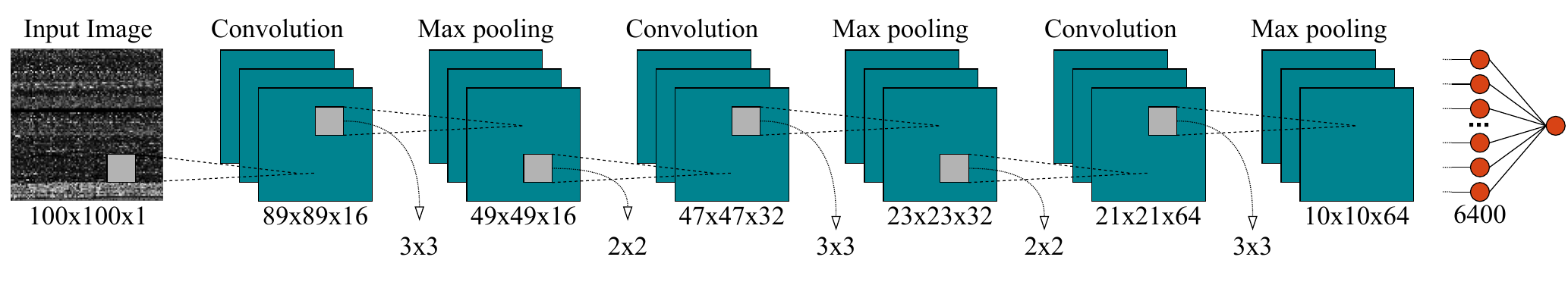}
    \caption{Architectural overview of the \textsc{Minos} classifier from Naseem \emph{et al.} \cite{minos}. The CNN contains three convolution layers, three pooling layers, and one fully connected layer. The input image shows a \texttt{Wasm} binary that is transformed into a grey-scale image.}
    \label{fig:minos-architecture} 
\end{figure}

%% file: figures/Training-Stats.tex
\begingroup
\setlength{\tabcolsep}{3pt} 
\renewcommand{\arraystretch}{1.1}
\begin{table}[b]
        \caption{Substitute network training evaluation after the last epoch for (a) one epoch (M-1) and (b) $50$ epochs (M-50).}
\begin{minipage}{.5\linewidth}
      \subcaption{M-1}
      \centering
\begin{tabular}{r|ccccc}
Fold            & 0 & 1 & 2 & 3 & 4 \\\hline
AUC             & 0.96   & 0.95   & 0.96   & 0.96   & 0.94   \\
Val. AUC        & 0.99   & 0.99   & 0.99   & 0.99   & 0.99   \\
Loss            & 0.29   & 0.30   & 0.29   & 0.29   & 0.34   \\
Val. Loss       & 0.13   & 0.14   & 0.15   & 0.15   & 0.14  
\end{tabular}
    \end{minipage}%
    \begin{minipage}{.5\linewidth}
      \centering
        \subcaption{M-50}
        
\begin{tabular}{r|ccccc}
Fold            & 0         & 1         & 2         & 3      & 4 \\\hline
AUC             & 1.00      & 1.00      & 1.00      & 1.00   & 1.00   \\
Val. AUC        & 1.00      & 1.00      & 1.00      & 1.00   & 1.00   \\
Loss            & 0.03      & 0.03      & 0.04      & 0.03   & 0.03   \\
Val. Loss       & 0.05      & 0.05      & 0.03      & 0.04   & 0.04  
\end{tabular}
    \end{minipage} 
        \label{tab:training}
\end{table}
\endgroup

%% file: figures/Section-Distribution.tex
\begin{figure}[t]
    \centering
    \includegraphics[width=1\textwidth]{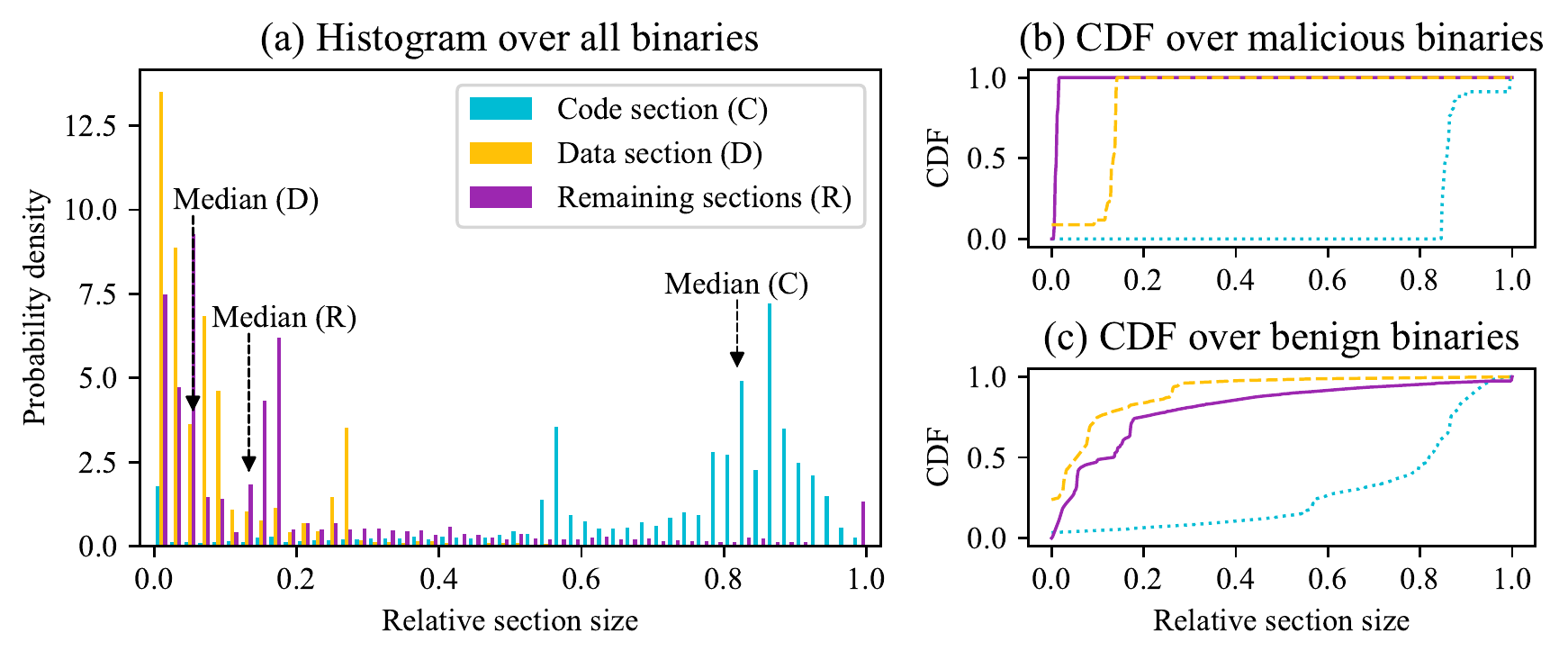}
    \caption{Histogram of relative section size (a) for code section, data section and all remaining sections for all binaries as described in Section \ref{sec:main:dataacquisition}. Cumulative density for relative section size for all malicious binaries (b) and benign binaries (c).}
    \label{fig:section-distribution}
\end{figure}

%% file: figures/Gadget.tex
\begin{figure}[t]
\centering
     \begin{subfigure}[t]{0.33\textwidth}
         \centering
         \begin{lstlisting}[style=wasm,escapechar=|,title= (a) Size-efficient gadget]
f64.const |\LstHighlight{.12}|0x0
drop
\end{lstlisting}
         \label{subfig:small-gadget}
     \end{subfigure}
     \hfill
     \begin{subfigure}[t]{0.62\textwidth}
         \centering
         \begin{lstlisting}[style=wasm,escapechar=|,title= (b) Binary representation of (a)]
0x44,|\LstHighlight{.65}|0x0,0x0,0x0,0x0,0x0,0x0,0x0,0x0
0x1a
\end{lstlisting}
         \label{subfig:small-gadget-bin}
     \end{subfigure}
     \hfill
     \newline
     \centering
     \begin{subfigure}[t]{0.33\textwidth}
         \centering
         \begin{lstlisting}[style=wasm,escapechar=|,title= (c) Optimizer-res. gadget]
(local $UID1 f64)
(loop $UID2
  local.get $UID1
  f64.const |\LstHighlight{.12}|0x0
  f64.add
  local.tee $UID1
  local.get $UID1
  f64.div
  f64.const 42
  f64.gt
  br_if $UID2
)
\end{lstlisting}
         \label{subfig:opt-gadget}
     \end{subfigure}
     \hfill
     \begin{subfigure}[t]{0.62\textwidth}
         \centering
         \begin{lstlisting}[style=wasm,escapechar=|,title= (d) Binary representation of (c)]
         |$\vspace{0pt}$|
0x03,0x40
0x20,0x02
0x44,|\LstHighlight{.65}|0x0,0x0,0x0,0x0,0x0,0x0,0x0,0x0
0xa0
0x22,0x02
0x20,0x02
0xa3
0x44,0x0,0x0,0x0,0x0,0x0,0x0,0x45,0x40
0x64
0x0d,0x00
0x0b
\end{lstlisting}
         \label{subfig:opt-gadget-bin}
     \end{subfigure}
     \hfill
    \caption{Size-efficient (a) and optimizer-resistant (c) gadget and their binary representation (b,d). Bytes that can be manipulated during the adversarial attack are highlighted in blue.}
    \label{fig:gadgets}
\end{figure}

%% file: chapter/Evaluation.tex
\section{Evaluation}
\label{sec:evaluation}
\subsection{Gadget Effectiveness}
\label{subsec:gadget_effectiveness}
\input{figures/Successrate-of-Gadget.tex}
Using our corpus of malicious samples (Section \ref{sec:main:dataacquisition}), we evaluate the effectiveness of our attack by creating adversarial examples for each binary. 
We consider the insertion density $d$ as the relative frequency of occurrence of our gadget, s.t. for a given density $d\in[0,1]$, for every 1000 instructions $d\cdot1000$ gadgets are added.
Fig. \ref{fig:eval:successrate} shows the misclassification rates of binaries with the size-efficient gadget (Fig. \ref{fig:eval:successrate}a) and the optimization-resistant gadget (Fig. \ref{fig:eval:successrate}b) against the \textsc{Minos} classifier \cite{minos} trained by Cabrera-Arteaga et \emph{al.} \cite{wasm-evade}. To the best of our knowledge, \textsc{Minos} is the only WebAssembly malware classifier that utilizes machine learning to classify malware directly on a representation of the binary itself.  To evaluate the effectiveness of our adversarial payloads at invoking misclassifications, we plot the misclassification rates for the original binary, the instrumented binary without adversarial payload and the adversarially crafted binaries. The original binaries are unaffected by the gadget density and never result in misclassification. For instrumented binaries without adversarial payloads, it becomes apparent that after a sufficiently large number of insertions, the classifier cannot detect the malicious binary even without the adversarial attack. Fig. \ref{fig:time-and-size}b shows the size increase of the binary through the addition of our gadgets. For each gadget, the  misclassification rates of the instrumented binaries start to increase significantly at a size of roughly $1.5\times$ the original binary. Considering that the larger the binary gets, the higher the compression rates and information loss are during downsampling, an increase in misclassification rates that correlates with a size increase can occur. Due to the difference in the number of added bytes per gadget, the misclassification rate for the larger optimization-resistant gadget increases at lower densities. 
However, for both gadgets, one can observe that the adversarially crafted binaries consistently outperform the binaries that are only instrumented, causing higher misclassification rates at lower densities. Additionally, adversarial payloads generated using the substitute models trained for one epoch consistently cause higher misclassifications at lower densities than payloads generated using the models trained for 50 epochs.
To further evaluate the misclassification caused by instrumenting the malicious binary, we additionally instrumented $50$ randomly selected benign binaries with the optimizer-resistant gadget that caused higher misclassification rates. At densities of both $0.1$ and $0.01$, the classifier correctly identified all evaluated benign binaries as benign, suggesting a tendency of the classifier to classify samples as benign. To evaluate the effectiveness of our method, we additionally generated adversarial payloads for the benign binaries that caused the substitute model to misclassify the binary as malicious. Using the substitute model trained for one epoch, we were able to successfully cause the target classifier to misclassify, on average, $77\%$ of the binaries over all folds at a density of $0.1$. 
Overall, at a density of $0.02$, both gadgets are shown to be successful in evading the target classifier for at least $70 \%$ of evaluated malicious binaries, while the misclassification rates for the instrumented binary without the adversarial payload are at or below $20 \%$, highlighting the effectiveness of our approach.
\subsection{Performance Analysis}
\label{subsec:exec-speed}
\input{figures/Time-and-Size-Overhead.tex}
To quantify the gadget's impact on the runtime of instrumented binaries, we measured the execution time in relation to the gadget density. This correlation is illustrated in Fig. \ref{fig:time-and-size}a.
We utilized a WebAssembly hashing library \cite{timing_benchmark_crypto} and performed $\num{5e5}$ rounds of SHA-256 hashing. A baseline was established by measuring the execution time without inserting the gadgets. The execution time of both gadgets is shown in relation to the baseline.
The insertion of the size-efficient gadget only results in a small constant increase in execution time, suggesting that the inserted gadget is not executed. WebAssembly is compiled using an ahead-of-time compiler, which includes optimization of the code. As the size-efficient gadget neither changes the data flow nor the control flow, the compiler likely identifies and removes those instructions during compilation. However, similar to \texttt{wasm-opt} \cite{wasm-opt}, this optimizer cannot detect the optimization-resistant gadget. As a result, the execution time increases linearly in the number of inserted gadgets. 
However, considering that a density of $0.02$ is enough to trick the target classifier, the increase in runtime is reasonable.

\begin{figure}[t]
    \centering
    \includegraphics[width=0.5\textwidth]{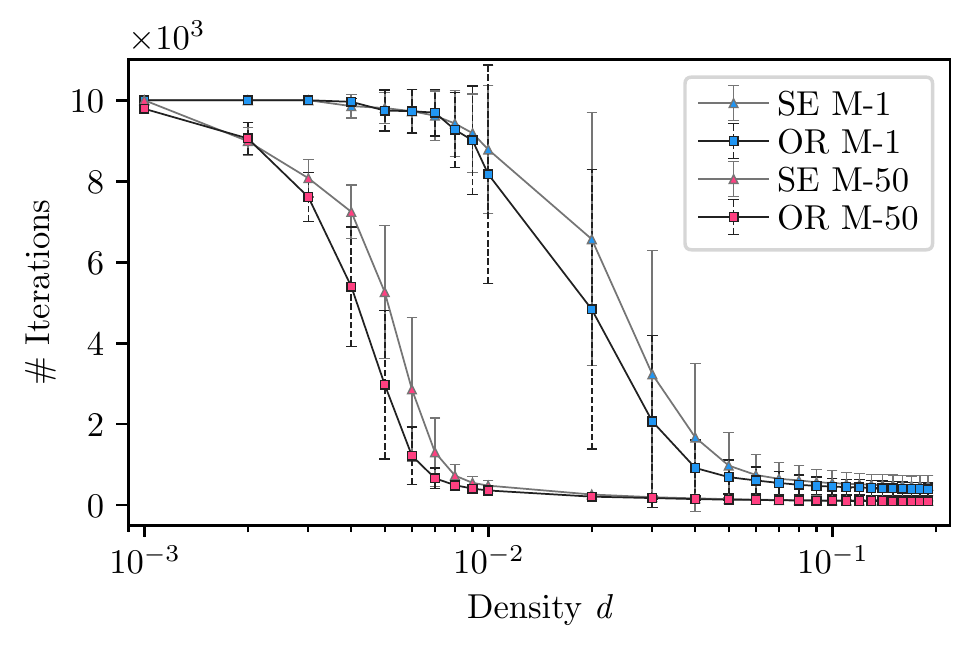}
    \caption{Average number of iterations ($y$-axis) required to achieve a confidence of $\num{1e-12}$ for a given gadget density ($x$-axis). Both the size-efficient gadget (SE) and the optimizer-resistant gadget (OR) are evaluated on the substitute model trained for one epoch (M-1) and $50$ epochs (M-50). The error bars show the standard deviation.}
    \label{fig:eval:adv-ex-iter-vs-density}
\end{figure}

Additionally, we evaluated the requirements for generating an adversarial example, which heavily depends on the gadget density. The number of iterations required to achieve a confidence of less than $\num{1e-13}$ within the shadow model was measured as a function of the chosen gadget density. The results are depicted in Fig. \ref{fig:eval:adv-ex-iter-vs-density}, which displays the average number of iterations required during the adversarial example generation over the applied gadget density. As both gadget types hold the same number of bytes utilized for the adversarial payload, they require a similar number of iterations to reach the confidence level.
The adversarial training optimization loop was run for a maximum of $\num{1e4}$ iterations. Overall, the lower the chosen density, the more iterations are required to reach the target confidence, as fewer bytes are available for adversarial crafting. While the adversarial examples crafted using the substitute model trained for one epoch outperform the adversarial examples crafted using the model trained for $50$ epochs, the adversarial example reaches the target confidence with fewer iterations on the model trained for $50$ epochs.
The execution time of a single iteration is 9.84 ms on an AMD Ryzen 9 7950X 16-Core Processor, which renders the attack feasible. Note that this optimization needs to only be performed once per malware.  However, an attacker could potentially exploit the low cost of generating new adversarial examples by regularly distributing new binaries to website visitors.

%% file: figures/Successrate-of-Gadget.tex
\begin{figure}[t]
    \centering
         \includegraphics[width=\textwidth]{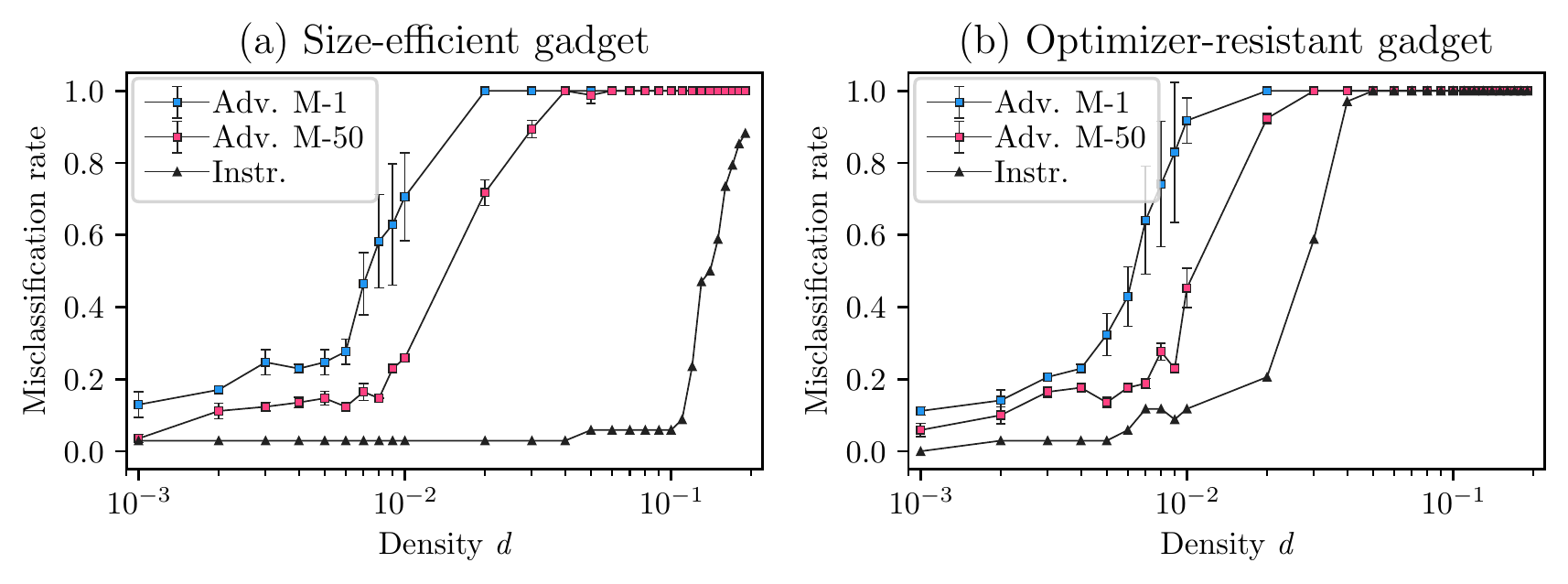} 
    \caption{\textsc{Minos} misclassification rate of binaries with size-efficient gadgets (a) and optimizer-resistant gadgets (b) against the pre-trained \textsc{Minos} \cite{minos} classifier by Cabrera-Arteaga \emph{et al.} \cite{wasm-evade}. Each plot depicts the misclassification rate of the original binary (Original), the instrumented binary \textit{without} adversarial payload (Instr.), and the misclassification rate of the binaries \textit{with} adversarial payload derived using Minos trained for one epoch (Adv. M-1) and for $50$ epochs (Adv. M-50). The adversarial misclassification rates are average over all five folds. The error bars depict the standard deviation.}
    \label{fig:eval:successrate}
\end{figure}

%% file: figures/Time-and-Size-Overhead.tex
\begin{figure}[t]
     \centering
         \includegraphics[width=\textwidth]{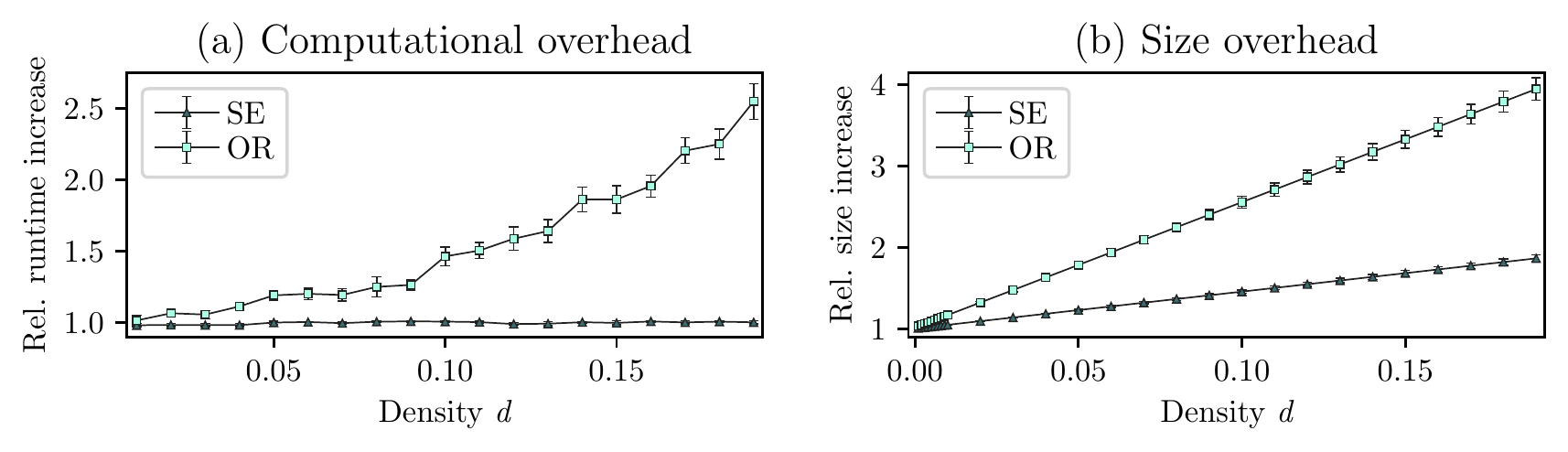} 
     \caption{Correlation between the insertion density and the relative increase in execution time (a) and size (b). Both the size-efficient gadget (SE) and the optimization-resistant gadget (OR) are evaluated. The $x$-axis represents the density of the gadgets, while the $y$-axis represents the relative execution time compared to the baseline (no gadget insertion) (a) and the relative increase of the binary's size in bytes (b). The average over the evaluated binaries is plotted, and the error bars represent the standard deviation.}
    \label{fig:time-and-size}
\end{figure}

%% file: chapter/Related-Work.tex
\section{Related work}
\label{sec:related-work}
The use of machine learning-based classifiers for detecting malware has been shown to be fast and effective in identifying binaries as malicious or benign. However, the robustness of these classifiers against adversarial inputs is often limited.
As more machine learning-based classifiers are utilized for detecting malware, malicious actors who want to distribute their malware have a high incentive to utilize evasion techniques to prevent detection. 
Especially for Windows Portable Executables (PEs), a number of classifiers and evasions exist. 
Existing adversarial evasions on classifiers that utilize a gray-scale image representation of the target binary \cite{13-COPYCAT-vis-eof-append} rely on FSGM \cite{explaining_advex} or Carlini \& Wagner \cite{Carlini_Wagner_Adv_example}, to generate a perturbation vector for the image \cite{12-vis-nonexec,13-COPYCAT-vis-eof-append,14-vis-nop-align}. 
However, in contrast to our attack, Liu \emph{et al.} \cite{12-vis-nonexec} directly apply the perturbation to the image representation of the binary. 
While they show a successful attack against the classifier, the generated adversarial example is not a valid binary anymore, rendering their evasion ineffective.
Khormali \emph{et al.} \cite{13-COPYCAT-vis-eof-append} generate the adversarial example and append the adversarial payload to the end of the file or at the end of a section. This ensures that the adversarial example is added into nonexecutable areas, and hence the original functionality remains. 
While this enables the addition of the adversarial payload into the malicious binary, a sophisticated defender can easily remove the payload by statically identifying unused bytes and masking them before classification, as they should have no impact on the classification performance.  
Using our attack methodology, the adversarial payload is placed inside the code section and directly baked into the control flow of the target binary, preventing a defender from easily removing the payload. Additionally, we have presented the optimization-resistant gadget that cannot be generally removed using an optimization pass.
Evasions against other network architectures that directly consider the sequence of bytes from Windows PE files generally insert adversarial payloads in unused bytes between sections \cite{7-raw-bytes-pe,8-slack-fgm,10-raw-mid-eof-injection}, in a new section \cite{7-raw-bytes-pe} or at the end of the file \cite{9-raw-cnn,10-raw-mid-eof-injection}. While these approaches generate executable binaries, it is rather easy to circumvent for a slightly more sophisticated detection model, e.g. one that first removes unused bytes or truncates sections or files. Either of our proposed gadgets is inserted directly into the instructions so that more sophisticated static analysis techniques, such as data flow and control flow analysis, are required to detect them fully. 
However, there are also numerous adversarial attacks against classifiers that classify a binary on more sophisticated features than just an image from its raw binary data, e.g. based on extracted features such as control flow, data flow, API calls, libraries, or dynamic features~\cite{20-api-arms-race,6-WinPE-SoK}. While the general procedure for generating the perturbation vector is similar, the application to the binary relies on transforming the target in a way that the corresponding features change. The interested reader is referred to Ling \emph{et al.} \cite{6-WinPE-SoK}, who provide an in-depth evaluation of different evasion techniques against Windows PE malware. 
Cabrera-Arteaga \emph{et al.} \cite{wasm-evade} proposed a malware evasion system against \texttt{Wasm} malware detectors and, in particular, \textsc{Minos}. However, their system relies on obfuscation to bypass detection frameworks, and they do not utilize adversarial attacks. 

%% file: chapter/Conclusion.tex
\section{Conclusion}
\label{sec:conclusion}
In this paper, we introduced a novel technique for placing adversarial payloads directly into the instruction stream using binary instrumentation to bypass machine learning-based malware detectors. We have demonstrated the effectiveness of our technique by crafting a grey-box adversarial attack against \textsc{Minos} \cite{minos}, a lightweight cryptojacking detection framework for WebAssembly presented at NDSS 2021. 
To place payloads inside the code section of the binary, we have introduced two semantic-preserving gadgets for \texttt{Wasm} binaries with a focus on size-efficiency and optimization-resistance, respectively.
We have collected an extensive dataset with both benign and malicious binaries by utilizing two existing benchmark datasets \cite{minesweeper,empirical-study} as well as results from a crawling campaign of one million websites from the Cisco Umbrella list \cite{cisco-1mil}. To populate this dataset, we used \texttt{wasm-mutate} \cite{wasm-mutate} to generate augmented binaries. Every sample was then assigned a label by querying the target model, i.e. \textsc{Minos} \cite{minos} provided by Cabrera-Arteaga \emph{et al.}\ \cite{wasm-evade}. 
All samples with their corresponding label were then used to train a substitute model of our targeted model. 
The challenge of creating a functional adversarial example inside a binary without altering the semantics was met by carefully inserting novel semantic-preserving gadgets. These gadgets can be injected freely into the code section of a \texttt{Wasm} binary without changing the semantics using binary instrumentation. 
Each gadget contains a number of bytes that carry the adversarial payload and can be manipulated freely during the attack phase. 
By attacking our substitute model, we successfully craft functional adversarial examples for cryptojacking binaries.
Using an insertion density of $0.02$ and the better-performing substitute network trained for one epoch (M-1), we are able to cause the target detector to misclassify all of the evaluated malicious binaries, demonstrating the effectiveness of our attack. Additionally, we show that our size-efficient gadget is removed during compilation resulting in only a negligible runtime overhead. The optimizer-resistant gadget, by design, is not removed before execution and thus leads to a linear overhead in the density. However, as a small insertion density of $0.02$ is sufficient in bypassing the classifier, the execution time is only increased by roughly $10\%$.
To prevent such attacks, we addressed typical countermeasures; However, as discussed by Tramèr \emph{et al.} \cite{DBLP:there_is_no_defense_against_adv_exmaples}, as long as the adversary is able to manipulate features used by a classifier, the threat of adversarial attacks cannot be fully mitigated. 
The success of our grey-box adversarial attack on \textsc{Minos} highlights the need for continued research and improvement of defences against adversarial attacks on machine learning-based malware detection frameworks.